\documentclass[12pt]{article}

\usepackage{graphicx}
\usepackage{bm}          

\begin{document}


\begin{center}
{\Large\bf \boldmath Effect of a Strong Laser on Spin Precession} 

\vspace*{6mm}
{Arsen Khvedelidze}\\      
{\small \it A.Razmadze Mathematical Institute, Tbilisi, GE-0193,
            Georgia\\      
            Joint Institute for Nuclear Research, Dubna, 141980,
            Russia }
\end{center}

\vspace*{6mm}

\begin{abstract}
The semiclassical dynamics of a  charged spin-1/2 particle in an
intense  electromagnetic plane wave is analyzed beyond the
electric dipole approximation and taking into account the leading
relativistic corrections to the Pauli equation. It is argued that
the adiabatic spin evolution driven by a low intensity radiation
changes its character drastically as an intensity of a laser is
increasing. Particularly, it  is shown that a charged particle
exposes a spin flip resonance at a certain pick value of a laser
field strength, which is determined by a particle's gyromagnetic
ratio.
\end{abstract}

\vspace*{6mm}

{\bf The problem  and result.\ }A good deal of a considerable
knowledge on a charged spinning particle interaction with a low
intensity laser has been gleaned from the extensive use of the
electric dipole approximation~\cite{Fedorov}. This approximation
works perfectly to describe the particle's classical trajectory as
well as to understand the adiabatic evolution of spin, represented
by the intrinsic angular momentum \cite{spincl}. With the growing
intensity of a radiation different relativistic corrections to the
charge motion become relevant \cite{SalaminHuHatsagortsyanKeitel},
\cite{MourouTajimaBulanov}. This demands to refuse the electric
dipole approximation and to take into account the influence of the
magnetic part of the Heaviside-Lorentz force. Entering to this
non-dipole region  a new physics become tangible. In this context,
the present talk aims to report on the manifestation of a such
non-dipole physics: \emph{a charged particle's spin flip resonance
induced by a strong laser field}.

It is arguably the best to describe the spin-flip resonance in the
so-called average rest frame, frame where the mean particle's
velocity vanishes. In this frame, as our calculations show, the
probability to flip for a spin, that is initially polarised along
the direction of propagation of the circularly  polarised
monochromatic plane wave, is given by an analog of the well-known
formula from the Rabi magnetic resonance problem \cite{Rabi}:
\begin{equation}\label{eq:spinoscill}
    \mathcal{P}_{\downarrow\uparrow}=
    \mathcal{A}_{\downarrow\uparrow}(\eta)\,\sin^2(\,\omega_S\,t)\,,
    \qquad
   \omega_{S}:= \frac{\omega_L|1-g|}{8}\,
   \sqrt{\,\kappa^2\eta^2 +(\eta^2-\eta^2_{\ast})^2\,}\,,
\end{equation}
The frequency $\omega_{S}\,$  differs from a laser circular
frequency $\omega_L\,$ and depends nonlinearly on a particle's
gyromagnetic ratio $g\,,$ as well as on a laser field strength
parameter~\cite{SarachikSchappert}:
\begin{equation}\label{eq:eta}
\eta^2=-2\,\frac{e^2}{m^2c^4}\,\langle\, A^2\rangle\,, \qquad
\langle \cdots \rangle  \-- \mbox{time~average}, \quad
 A \-- \mbox{a laser~gauge~potential\,.}
\end{equation}
The flipping amplitude $\mathcal{A}_{\uparrow\downarrow}(\eta)$ in
(\ref{eq:spinoscill}) has the following resonance form
\begin{equation}\label{eq:spinflipamplit}
   \mathcal{A}_{\downarrow\uparrow}(\eta) = \frac{\kappa^2\eta^2}{\kappa^2\eta^2+
    (\eta^2-\eta^2_{\ast})^2}\,, \qquad
    \kappa^2
:= \frac{2\,g^2}{(1-g)^2}\,,\qquad \eta^2_{\ast}: =\frac{4}{g-1}\,.
\end{equation}

{\bf Sketch of the calculations.\ }To get the above results the
recently elaborated  method \cite{PaulArsen08} is extended from
the classical to the quantum case.  The conventional semiclassical
approach attitude, when a charged particle motion in a given
electromagnetic background is studied classically within the
non-relativistic Hamilton-Jacobi theory, is adopted. At the same
time, the spin evolution is treated quantum mechanically, as
required by the spin nature, using the Pauli equations with the
leading relativistic corrections. The spin-radiation interaction
is encoded in the effective spatially homogeneous magnetic field
configuration, which is determined by the geometry of a particle's
classical trajectory. The described approximation is formulated
mathematically as follows. The laser radiation is modelled by the
\textit{elliptically polarized monochromatic plane wave}
propagated along the $z$-axis
\begin{equation}\label{eq:gpot}
    A^\mu : = a\biggl(0\,, \ \varepsilon\cos(\omega_L\xi)\,,\
    \sqrt{1-\varepsilon^2}\sin(\omega_L\xi)\,,\ 0
    \biggl)\,,\qquad  \xi = t
-\frac{z}{c}\,,
\end{equation}
where  $0\leq\varepsilon\leq 1$ is the light polarisation parameter,
and the constant ${a}$ is related to the laser field strength
(\ref{eq:eta}) , \(\ \eta^2={e^2a^2}/{m^2c^4}\,. \) A charged
spin-1/2 particle is in a pure quantum state $\Psi$ admitting the
semiclassical charge $\& $ spin decomposition,
\begin{equation}\label{eq:spin-charge.decomp}
|\Psi\rangle = \sum_{i=0,1}\,\sum_{\alpha=\pm}\ c_{\alpha,
i}\,|\psi_\alpha\rangle\otimes|\chi_i\rangle\,.
\end{equation}
Two states, $|\psi_\pm\rangle\,,$ \footnote{To simplify
expressions, the initial state is assumed to have  only one
nonzero coefficient, $c_{+,0}\,.$ Note that, the unit
normalization condition on the WKB wave function fixes this
coefficient $\pi c^2_{+,0} = 2m\,\omega_\mathrm{P}\,,$
$\omega_\mathrm{p}$ \---  a particle's fundamental frequency.} \
are the linearly independent WKB solutions to the Schr\"{o}dinger
equation for a charged spinless particle moving in the background
(\ref{eq:gpot}). According to the semiclassical calculations  the
spin state vectors $|\chi\rangle$ in the decomposition
(\ref{eq:spin-charge.decomp}) satisfy the spin evolution equation:
\begin{equation}\label{eq:spin.evol}
i\frac{\mathrm{d}}{\mathrm{d}t}\, |\chi \rangle = -\,
\frac{ge}{4mc}\,
  \mathbf{B}^\prime(t)\cdot \boldsymbol{\sigma}\,
|\chi\rangle\,,
\end{equation}
where the spatially homogeneous magnetic field
$\mathbf{B}^\prime(t)\,$ is accounted for a  laser filed coupling to
a spin moving in the laboratory  frame  with the velocity $
\boldsymbol{v}$ and acceleration $\, \boldsymbol{a}\,,$
\begin{equation}\label{eq:eff.mag}
   \mathbf{B}^\prime(t) :=
\left(\boldsymbol{B}- \frac{1}{c}\, \boldsymbol{v}\times
\boldsymbol{E}\right)\, + \, {\frac{m}{egc}\, \boldsymbol{v}\times
\boldsymbol{a}}\,.
\end{equation}
Here the term in parenthesis is magnetic field seen in the
particle's instantaneous rest frame and evaluated along the
particle's classical orbit. The last contribution in
(\ref{eq:eff.mag}) corresponds to the leading part of the
so-called Thomas precession correction due to the non-vanishing
curvature of a particle's trajectory \cite{spincl}.

Analysis of the equations begins with the derivation of the exact
solution to the classical non-relativistic Hamilton-Jacobi equation
for spinless particle moving in the electromagnetic background
(\ref{eq:gpot}). Solving this Hamilton-Jacobi problem
\cite{PaulArsen08} one determine both, the WKB solution to the
Schr\"{o}dinger equation and the effective magnetic field
(\ref{eq:eff.mag}). We assume below that the particle's classical
trajectory $\boldsymbol{x}(t)\,$ fulfills the initial condition
$\boldsymbol{x}(0)=0\,$  and fix also the frame, where the average
of a particle's velocity component, orthogonal to the wave
propagation direction vanishes, \(\
<\boldsymbol{\upsilon}_\bot>=0\,. \) To find the effective magnetic
field we owe from \cite{PaulArsen08} the expression for a particle's
velocity
\begin{equation}\label{eq:xvelocity}
\boldsymbol{\upsilon}=
\left[-c\eta\varepsilon\,\mathrm{cn}\left(\omega_{L}^\prime
t,\,\mu\right), \,
 -c\eta \sqrt{1-\varepsilon^2}\,
\mathrm{sn}\left(\omega_{L}^\prime t, \,\mu\right),\,
c-c(1-\beta_z)\,
    \mathrm{dn}\left(\omega_{L}^\prime t,\,
    \mu\right)
\right]\,,
\end{equation}
as well as the expression for its $z$-coordinate: \(z(t) = c t -
\frac{c}{\omega_L}\,\mathrm{am}(\omega_L^\prime t,\, \mu)\,. \)
The argument of the elliptic Jacobian functions and the Jacobian
amplitude function, $\mathrm{am}(u,\, \mu)\,,$  is the laboratory
frame time $ t\,,$ scaled by the non-relativistically Doppler
shifted laser frequency \(\ \omega_L^\prime= \gamma_z\omega_L\,,\
\gamma_z=1-\upsilon_z(0)/c\,\) and the modulus $\mu$ is
\(\gamma^2_z \mu^2=\eta^2 (1-2\,\varepsilon^2)\,. \)\footnote{The
expression (\ref{eq:xvelocity}) is the correct one only for $\mu^2
\in (0,1)\,.$ The solution outside this interval can be
reconstructed exploiting the modular properties of the Jacobian
function. See details in ref. \cite{PaulArsen08}.} Using this
solution the exact expression for the effective magnetic field
$\mathbf{B}^\prime(t)\,$  can be found:
\begin{eqnarray*}\label{eq:eff}
 \mathbf{B}^\prime=\frac{a\omega_L^\prime}{gc}
 \left(\sqrt{1-\varepsilon^2}\left[(g+1)\mathrm{dn}-\gamma_z\right]
\mathrm{cn},
\varepsilon\left[(g+1)\mathrm{dn}-\gamma_z(1-\mu^2)\right]\mathrm{sn},
\varepsilon\,
\sqrt{1-\varepsilon^2}\left[
\mathrm{dn}-g\gamma_z^{-1}\right]\right).
\end{eqnarray*}

\textit{The resonant oscillations.\---} In the average rest frame,
$<\boldsymbol{\upsilon} >=0\,,$ when laser beam is circularly
polarised, $\varepsilon^2=1/2\,,$  the expression for the
effective magnetic field $\mathbf{B}^\prime$ simplifies to the
constant magnitude field:
\begin{equation}\label{eq:rotmag}
\mathbf{B}^\prime (t)= |\mathbf{B}^\prime |\,\boldsymbol{n}(t)\, ,
\qquad |\mathbf{B}^\prime |:=
\frac{a\omega_L|1-g|}{2gc}\,\sqrt{\kappa^2+ \eta^2}\,,
\end{equation}
aligned the unit time dependent vector $\boldsymbol{n}(t):=
\left(\sin\theta\cos\omega_L t, \ \sin\theta\sin\omega_L t, \
\cos\theta\right).$ The effective magnetic field (\ref{eq:rotmag})
rotates with the frequency $\omega_L$ about the axis inclined with
respect to the field.  The inclination  angle $\theta$ is
determined from the relation:   \(\ \eta\tan\theta =\kappa\,. \)

Therefore,  \emph{the effective laser-spin interaction for the
circular polarised radiation is precisely the  famous rotated
magnetic field describing the nuclear magnetic resonance
phenomenon~!} Having in mind this observation one can use the
well-known exact Rabi solution \cite{Rabi} to find the semiclassical
evolution of spin-1/2 particle. Particularly, a straightforward
calculations lead to the expressions (\ref{eq:spinoscill}) and
(\ref{eq:spinflipamplit}) for the spin flipping probability
announced at the beginning of this report.

{\bf Conclusion.} In the present  talk  the non-dipole effect of a
strong laser on the spin of a charged particle was described quantum
mechanically, while the evolution of position and momentum of a
particle itself were treated according to the classical Newton
equations with  complete Heaviside-Lorentz force. The derived
results indicate a very different spin physics in a high intensity
laser field versus to a low intensity adiabatic regime.

\textit{Acknowledgment.\---}The research was supported in part by
the Georgian National Science Foundation under Grant No.
GNSF/ST06/4-050, the Russian Foundation for Basic Research Grant No.
08-01-00660  and by the Ministry of Education and Science of the
Russian Federation Grant No. 1027.2008.2.



\begin{thebibliography}{99}\itemsep -1mm
\bibitem{Fedorov}M.V.~Fedorov, \textit{``Electron in a Strong Optical Field''},
Nauka, Moscow,  1991.
\bibitem{spincl}J.~Frenkel, Z.\ Phys.\ {\bf 37}, 243 (1926);
L.H.~Thomas, Nature\  {\bf 117}, 514 (1926).
\bibitem{SalaminHuHatsagortsyanKeitel}
Y.I.~Salamin, {\em et al}., Phys.\ Rep.\ {\bf 427}, 41 (2006).
\bibitem{MourouTajimaBulanov}G.A.~Mourou,
T.~Tajima and S.V.~Bulanov, Rev.\ Mod.\ Phys.\ {\bf 78}, 309 (2006).
\bibitem{Rabi}I.I.~Rabi,
Phys.\ Rev.\ {\bf 51}, 652 (1937).
\bibitem{SarachikSchappert}E.S.~Sarachik and G.T.~Schappert, Phys.\ Rev. {\bf D1}, 2738
(1970).
\bibitem{PaulArsen08}P.~Jameson and A.~Khvedelidze,
Phys.\ Rev. {\bf A77}, 053403 (2008).
\end{thebibliography}
\end{document}